\documentclass[a4paper,11pt]{article}
\pdfoutput=1 
\usepackage{jinstpub} 
\usepackage{lineno}
\title{\boldmath Intrinsic time resolution of 3D-trench silicon pixels for charged particle detection }

\author[a]{L.~Anderlini,}
\author[b]{M.~Aresti,}
 \author[a,h]{A.~Bizzeti,}
 \author[c,d]{M.~Boscardin,}
 \author[b,1]{A.~Cardini\note{Corresponding author}}
\author[d,i]{G.-F.~Dalla Betta,}
\author[e,j]{M.~Ferrero,}
\author[d,i]{G.~Forcolin,}
\author[b,k]{M.~Garau,}
\author[b]{A.~Lai,}
\author[b,k]{A.~Lampis,}
\author[b]{A.~Loi,}
\author[a,l]{C.~Lucarelli,}
\author[c]{R.~Mendicino,}
\author[f,m]{R.~Mulargia,}
\author[e,n]{M.~Obertino,}
\author[f]{E.~Robutti,}
\author[c]{S.~Ronchin,}
\author[e,o]{M.~Ruspa,}
\author[g]{S.~Vecchi}
\affiliation[a]{INFN, Sezione di Firenze, Firenze, Italy}
\affiliation[b]{INFN, Sezione di Cagliari, Cagliari, Italy}
\affiliation[c]{FBK, Fondazione Bruno Kessler, Trento, Italy}
\affiliation[d]{INFN, TIFPA, Trento, Italy}
\affiliation[e]{INFN, Sezione di Torino, Torino, Italy}
\affiliation[f]{INFN, Sezione di Genova, Genova, Italy}
\affiliation[g]{INFN, Sezione di Ferrara, Ferrara, Italy}
\affiliation[h]{Dipartimento di Scienze Fisiche, Informatiche e Matematiche dell'Universit\`a di Modena e Reggio Emilia, Modena, Italy}
\affiliation[i]{Dipartimento di Ingegneria Industriale, Universit\`a di Trento, Trento, Italy}
\affiliation[j]{Dipartimento di Scienze del Farmaco, Universit\`a del Piemonte Orientale, Novara, Italy}
\affiliation[k]{Dipartimento di Fisica dell'Universit\`a di Cagliari, Cagliari, Italy}
\affiliation[l]{Dipartimento di Fisica e Astronomia dell'Universit\`a degli Studi di Firenze, Sesto Fiorentino, Italy}
\affiliation[m]{Dipartimento di Fisica dell'Universit\`a di Genova, Genova, Italy}
\affiliation[n]{Dipartimento di Scienze Agrarie, Forestali ed Alimentari dell'Universit\`a di Torino, Grugliasco, Italy}
\affiliation[o]{Dipartimento di Scienze della Salute, Universit\`a del Piemonte Orientale, Novara, Italy}

\emailAdd{alessandro.cardini@cern.ch}

\abstract{
In the last years, high-resolution time tagging has emerged as a promising tool to tackle the problem of high-track density in the detectors of the next generation of experiments at particle colliders. Time resolutions below 50~ps and event average repetition rates of tens of MHz on sensor pixels having a pitch of 50~$\mu$m are typical minimum requirements. This poses an important scientific and technological challenge on the development of particle sensors and processing electronics. 
The TIMESPOT initiative (which stands for TIME and SPace real-time Operating Tracker) aims at the development of a full prototype detection system suitable for the particle trackers of the next-to-come particle physics experiments. This paper describes the results obtained on the first batch of TIMESPOT silicon sensors, based on a novel 3D MEMS (micro electro-mechanical systems) design. We demonstrate that following this approach, the performance of other ongoing silicon sensor developments can be matched and overcome. In addition, 3D technology has already been proved to be robust against radiation damage. A time resolution of the order of 20~ps has been measured at room temperature suggesting also possible improvements after further optimisations of the front-end electronics processing stage.
}

\arxivnumber{2004.10881v2} 

\proceeding{Accepted for publication on Journal of Instrumentation}
\begin{document}
\maketitle
\flushbottom

\section{Introduction}

To pursue the quest for new physics phenomena and to constrain our knowledge of the Standard Model of Particle Physics with increased precision, CERN will upgrade the Large Hadron Collider (LHC) in the near future~\cite{HiLumiLHC}. Such upgrade (High Luminosity LHC, HL-LHC) will allow the experiments to collect data sets ten times larger than in the current conditions. Moreover, post-LHC colliders are presently under consideration by the scientific community~\cite{FCC,CEPC}.
Future machines will be pushed to their technological limits in order to increase the centre-of-mass energy and collision rate. This will impose severe requirements on new-generation detectors operating in environments with unprecedented event pile-up and radiation levels.
 The increased particle flux will dramatically enhance the irradiation of all the detectors; specifically, for the innermost layers of the future tracking systems, the foreseen fluence ranges between $10^{16}$~${\rm n}_{\rm eq}$/cm$^2$ (1~MeV neutron equivalent/cm$^2$) at HL-LHC and up to $10^{17}$~${\rm n}_{\rm eq}$/cm$^2$ per year at post-LHC colliders, thus reducing, when not spoiling, their performance and shortening their lifetimes. 
The increased number of simultaneous interactions at each beam crossing (between 100 and 200 at HL-LHC, and up to ten times more at the next generation of hadron colliders) will make it difficult to associate the detected particles to the parent collision. In this scenario, the time information for each charged track can crucially improve the event reconstruction: timing detectors are an option under discussion or already a design choice, referred to as \emph{timing layer}~\cite{CMS-TL,ATLAS-TL}.
For experiments at colliders of the next-to-come generations (\mbox{e.g.} LHCb Upgrade 2~\cite{LHCb-PII-Physics}, FCC~\cite{FCC}), the combination of time and space measurements in a single device is needed. Depending on the experiment and on the detector considered, a single-hit measurement with a time resolution ranging from 10~ps to 100~ps appears to be adequate~\cite{RiassuntoEUStrat}.

Nowadays silicon sensors are the only detectors able to provide excellent spatial resolution, fine pixel size and large area coverage in harsh radiation environments. Different technologies are in use and relevant progress has been achieved in the last decade to tackle the challenges discussed above. Several R\&D studies are ongoing to improve the timing performances of silicon sensors towards the frontier of radiation-hard tracking systems with unprecedented space and time resolution. 

Small capacitance, high signal-to-noise ratio, speed and spatial response uniformity are key ingredients to be considered when designing  a high-resolution timing detector. To meet these requirements, thin silicon sensors with a gain layer have been recently introduced. Ultra Fast Silicon Detectors (UFSD), based on Low Gain Avalanche Diode (LGAD) technology, presently achieve 30~ps resolution up to fluences of 1$-$2$\times 10^{15}$~${\rm n}_{\rm eq}$/cm$^2$~\cite{ufsd}. Such devices are currently the baseline for the forward part of the ATLAS and CMS timing layers at HL-LHC~\cite{CMS-TL,ATLAS-TL}.
Sensors with three-dimensional electrodes (3D sensors) are also a valid alternative. Since their introduction in 1997~\cite{3D}, this technology has consolidated: 3D detectors are presently used at the LHC experiments in regions very close to the beam (CMS-PPS~\cite{pps}, ATLAS-IBL~\cite{ibl}) and are good candidates for the HL-LHC tracking detector upgrades. The sensors are characterised by cylindrical electrodes penetrating deep into the bulk material, perpendicularly to the surface.  This unique structure, which decouples the charge carrier drift distance from the sensor thickness, exhibits very good radiation hardness, probed up to fluence of $3\times 10^{16}$~${\rm n}_{\rm eq}$/cm$^2$~\cite{3Dradhard}.  
The geometry of 3D sensors is also beneficial in terms of timing performance. Very short collection times can be achieved by choosing small inter-electrode spacing without reducing the substrate thickness, thus preserving the signal amplitude. The fact that the charge carriers are collected perpendicularly to the sensor thickness minimises time uncertainties due to nonuniform ionisation density (delta rays) and charge carriers diffusion. 

In this paper a novel 3D sensor design based on parallel trench electrodes, optimised to reduce electric field non¢uniformities, is presented. The time resolution of this device has been measured and characterised with a charged particle beam, and results are discussed in the following. 

\section{Sensors}
\label{sec:sensor}

\subsection{Sensor design}
In 3D sensors the amount of charge deposit and, consequently, the signal amplitude are independent of the inter-electrode distance and electrode shape. This peculiarity allows a large freedom in the choice of the sensitive volume geometry definition and, at the same time, makes its design of crucial importance in obtaining optimal performance. The development of a 3D silicon sensor with enhanced timing capabilities starts therefore from a detailed optimisation of its layout. 

\begin{figure}[t!]
    \centering
    \includegraphics[width=0.95\textwidth]{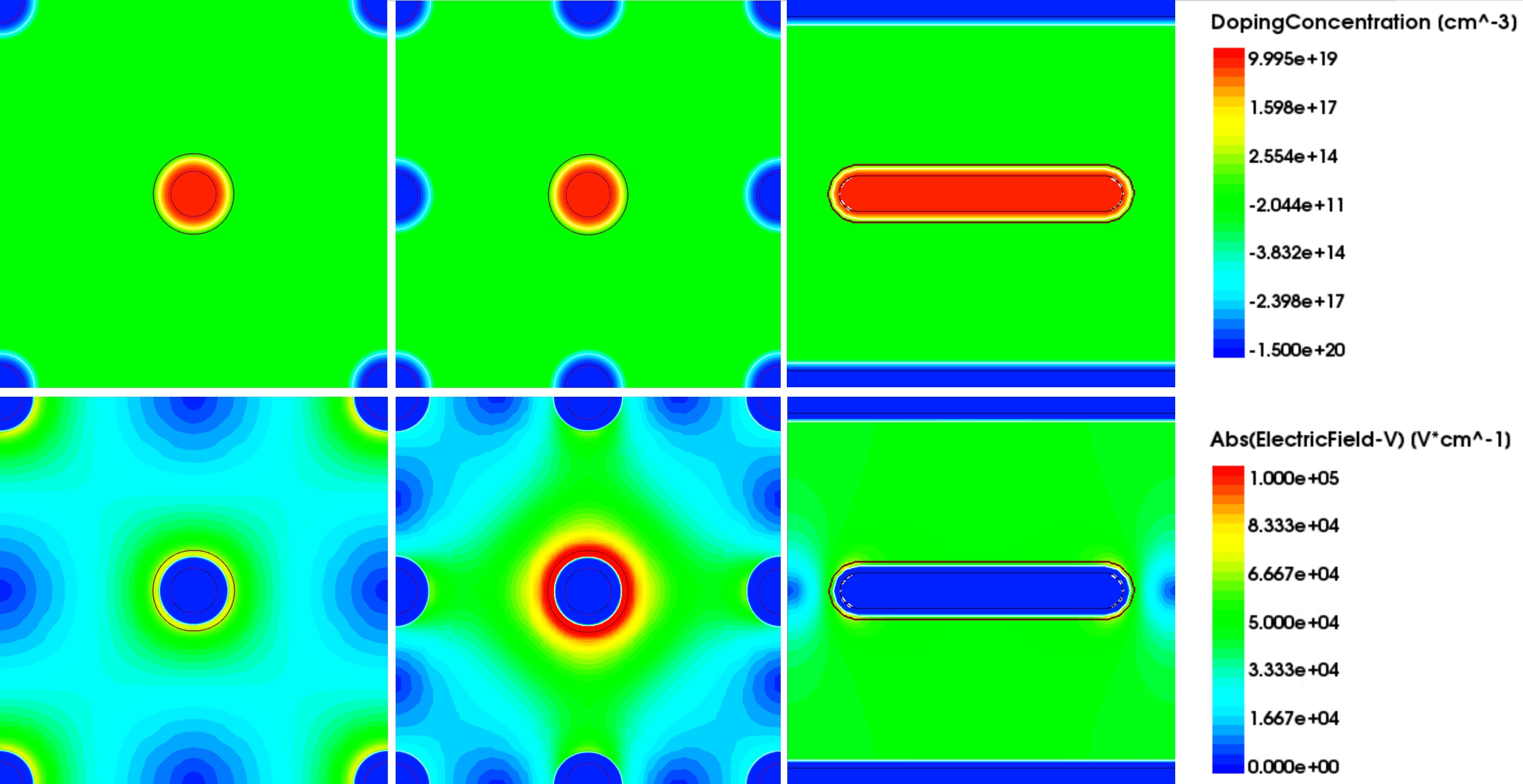}
    \caption{TCAD 2D model simulation of three different electrode geometries at bias voltage $V_{\rm bias} = 100$~V. The upper maps show the electrode geometries and doping profiles. The lower maps present the generated electric field inside the sensible area. From left to right: 3D-column (five electrodes), 3D-column (nine electrodes), 3D-trench. The 3D-trench geometry presents by far the most uniform electric field.}
    \label{fig:geomComp}
\end{figure}
At a first-order simplified analysis, the time resolution of a system for particle detection (sensor and its read-out electronics) is:
%
\begin{equation}\label{eq1}
   \sigma_{\rm t} = \sqrt {\sigma^2_{\rm ej} + \sigma^2_{\rm tw} + \sigma^2_{\rm dr} + \sigma^2_{\rm TDC} + \sigma^2_{\rm un}} \, ,
\end{equation}
where $\sigma_{\rm ej}$ is the electronic jitter, which depends on the front-end electronics rise time and signal-to-noise ratio; $\sigma_{\rm tw}$ depends on fluctuations of the signal amplitude, which cannot be minimised by design but only by dedicated signal processing; 
$\sigma_{\rm dr}$ depends on the effect of longitudinally disuniformities in the energy deposit due to delta rays, which is negligible in 3D detectors because the charge carriers are collected perpendicularly to the direction of ionizing particles crossing the sensor;
$\sigma_{\rm TDC}$ depends on the digital resolution of the electronics.
The $\sigma_{\rm un}$ contribution corresponds to the time dispersion caused by unevenness in the signal shapes, which are due to the different possible drift paths of the charge carriers in the sensor. The geometric characteristics of the sensor partially affect the $\sigma_{\rm ej}$ term, because noise depends linearly on sensor capacitance. On the other hand, the $\sigma_{\rm un}$ term depends only on the geometry of the sensitive volume. As established by the Shockley-Ramo theorem~\cite{Ramo} the signal is determined by the instantaneous current, $i$, induced at the electrodes by the charge carriers moving along their drift paths, through contributions of the form 
\begin{equation}\label{eq2}
i = q \, \textbf{E}_w \cdot \textbf{v}_d \, ,
\end{equation}
where $\textbf{E}_w$ is the weighting field and $\textbf{v}_d$ is the carrier's drift velocity. In order to minimise the $\sigma_{\rm un}$ term, maximum uniformity in the electric field must be obtained by design. 
Concerning the $\textbf{v}_d$ term, instead, it is important to reach the velocity saturation regime at a relatively low value of the biasing voltage, which is an additional favourable consequence of the short inter-electrode distance in any 3D device.
Figure~\ref{fig:geomComp} shows a comparison among the electric field maps of different geometrical solutions (3D-column and 3D-trench) obtained with the TCAD simulation package~\cite{TCAD}. Independently of the column size and the pixel pitch, the 3D-column geometries exhibit relatively large weak-field areas, responsible of inefficiencies and smaller induced current signals. 
This effect is considerably reduced in the 3D-trench geometry, which gives maximum field uniformity. 
Accurate simulations have extensively verified the enhanced timing performance of 3D-trench geometry with respect to the 3D-column ones, as illustrated in the next section.

\subsection{Modelling and performance simulation of 3D Sensors}\label{sec:simulation}  

Owing to their structure based on vertical electrodes, 3D silicon sensors allow a large degree of customisation of the electrode geometry according to the application. For fast timing applications in high luminosity environment, it is straightforward to develop a 3D silicon sensor with a small inter-electrode distance. The advantage of smaller electrode distances is a greater electric field achieved at lower bias voltages, which minimises the charge collection time and at the same time allows operating the sensor in carrier saturation velocity regime. A smaller charge collection time also reduces the probability of carrier trapping by radiation-generated defects, which favours the radiation hardness of the device. Inter-electrode distance is however limited by pixel capacitance, which should be minimised, and by the physical size and pitch of the pixel, which needs to be matched to a complete front-end channel for signal readout and processing. In this respect, pixels have a typical pitch of about 50~$\mu$m.

Besides short collection times, fast timing is favoured by a uniform shape distribution of the induced signals. This is obtainable by customising the geometry of both the electric and the carrier velocity fields within the sensor sensitive volume.

Several different layouts based on hexagonal and square pixel shapes with different electrode geometries (column and trench) were modelled and compared. A wide and detailed analysis of the different geometries can be found in ref.~\cite{PhD_ALoi}. Among all the explored configurations, the most promising design is obtained with a parallel trench geometry.

\begin{figure}[t!]
    \centering
    \includegraphics[width=1.0\textwidth]{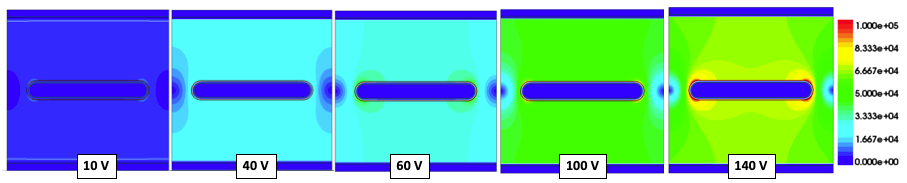}\\
    \centering
    \includegraphics[width=1.0\textwidth]{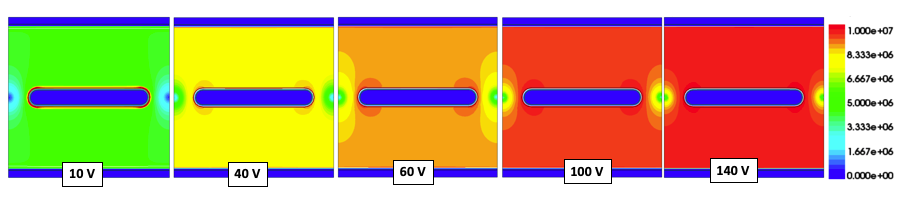}
    \caption{(Top) Electric field magnitude maps simulated for different bias voltages. The units in colour scale are V/cm.
    (Bottom) Electron drift velocity maps for different bias voltages. The units in colour scale are cm/s. Both plots are obtained with TCAD simulations.}
    \label{fig:EFieldeVel}
\end{figure}

Figure~\ref{fig:EFieldeVel} top shows the electric field maps of the parallel trench configuration, while figure~\ref{fig:EFieldeVel} bottom shows the drift velocity maps for electrons, both obtained with TCAD simulations. An electric field above $10$~kV/cm, corresponding to the beginning of the saturation velocity regime, is obtained at bias voltages of a few tens of Volts. Moreover, the low field regions, having a slower sensor response, are reduced to a small fraction of the total volume already at around 100~V.

\begin{figure}[t!]
    \centering
    \includegraphics[width=1.0\textwidth]{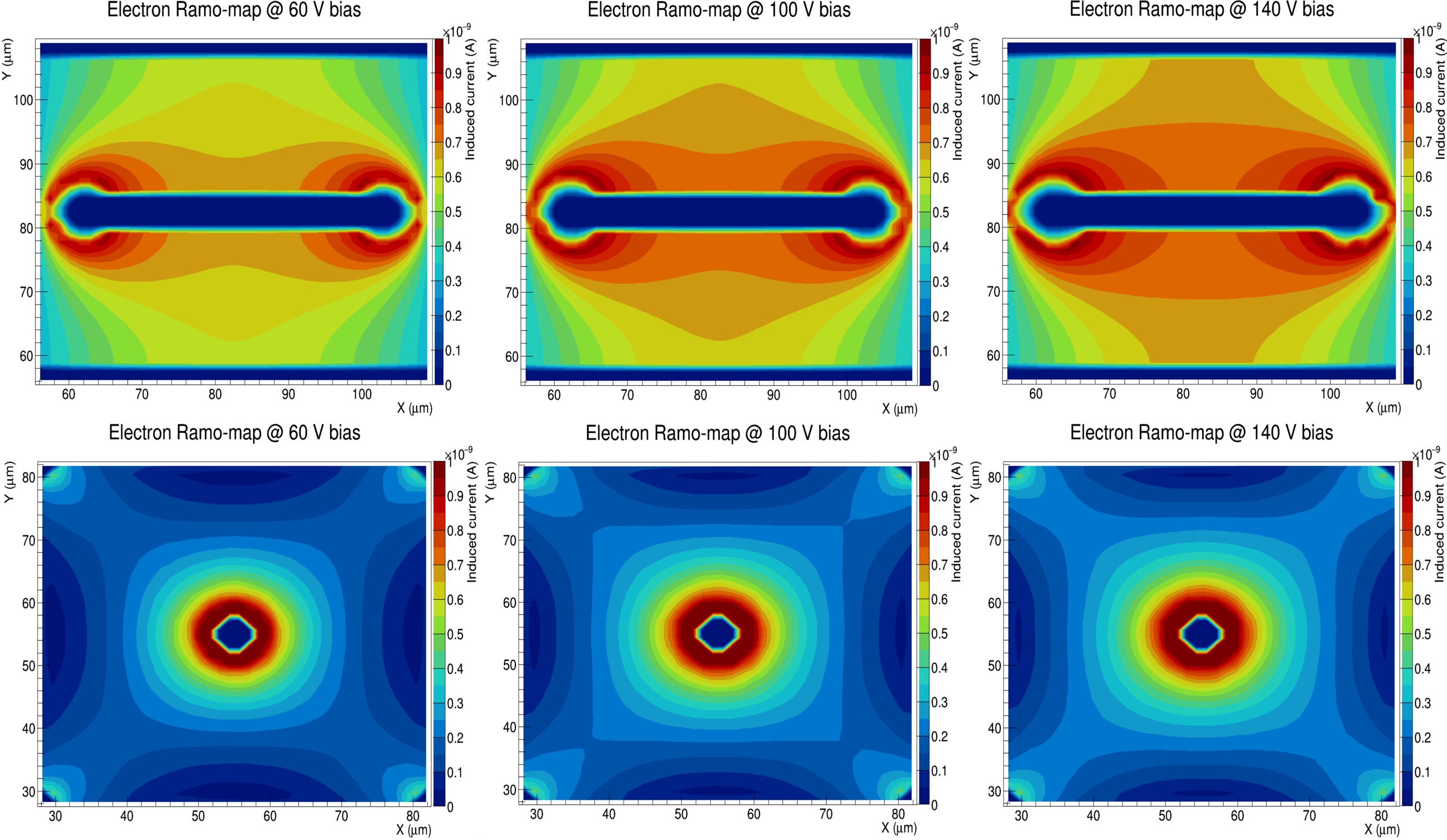}
    \caption{Ramo maps for electrons at different bias voltages: (top) trench geometry and (bottom) five-column geometry. Plots are obtained from TCAD simulations.}
    \label{fig:RamoMaps}
\end{figure}

A second analysis based on a quasi-stationary simulation named \emph{Ramo maps} was used to evaluate the induced current given by a charge moving in each point of the sensitive volume between the electrodes. 
The Ramo map (figure~\ref{fig:RamoMaps}) represents the contribution to the induced current $i(t)$ of each space point according to eq.~\ref{eq2}.
The advantage of the trench geometry with respect to the column geometry is shown in figure~\ref{fig:RamoMaps} by comparing the Ramo maps obtained in the two cases. 
It can be noticed that large contributions to the induced current are given only in the proximity  of the collecting electrode for the column case, while extend to almost the complete volume in the trench case.

The designed 3D sensor model was then used to estimate charge collection time, simulating charge deposits from ${\cal O}(10^3)$ MIPs perpendicularly impinging on the sensor surface at random positions. The simulations were performed using a GEANT4-based Montecarlo~\cite{Geant4,Geant4_b} to obtain an accurate description of the energy deposit in the sensor volume ($dE/dx$).
Static sensor properties (electric field, weighting field and velocity maps) were calculated with the Synopsys Sentaurus TCAD package~\cite{TCAD}. The carrier dynamics was entirely simulated by means of the TCoDe software~\cite{TCode}, developed within the TIMESPOT collaboration to increase processing speed of induced signal calculations. 

\begin{figure}[t!]
    \centering
    \includegraphics[width=1.0\textwidth]{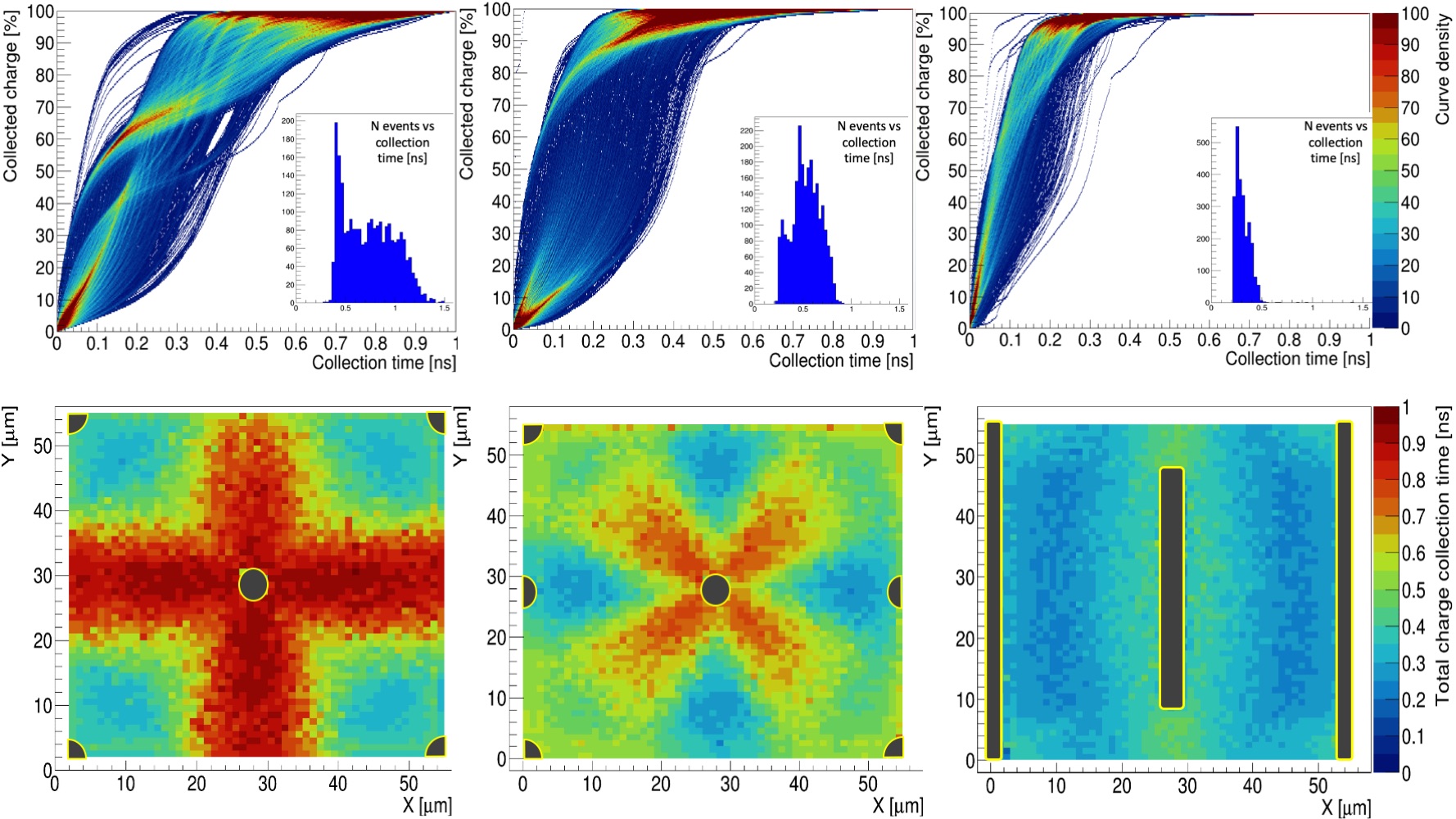}
    \caption{Time performance comparison among three different 3D geometries at $V_{\rm bias} = 100$~V (from left to right: five columns, nine columns and trench geometry). (Top) percentage of total charge collected on the electrodes versus time. (Top inserts) distribution of charge collection time for the three geometries. (Bottom) time for complete charge collection versus impact point for the same geometries. Each simulation is based on about 3\,000 MIP tracks. 
    }
    \label{fig:3Dcomparison}
\end{figure}

An important result from such simulations is shown in figure~\ref{fig:3Dcomparison}, where the charge collection time for different geometries (trench and column) are compared. 
It can be observed that going from left to right the charge collection times become shorter and consequently their distributions become narrow and with a single-peaking structure, which leads to an improved response uniformity. In the optimal case, the 3D-trench, the time distribution is peaked and appears as a Gaussian with a tail at larger arrival times, due to the slightly slower areas in proximity to the collecting trench (figure~\ref{fig:3Dcomparison} bottom right). 

\subsection{Fabrication of 3D-trench sensors}  

\begin{figure}[h!]
    \centering
    \includegraphics[width=0.95\textwidth]{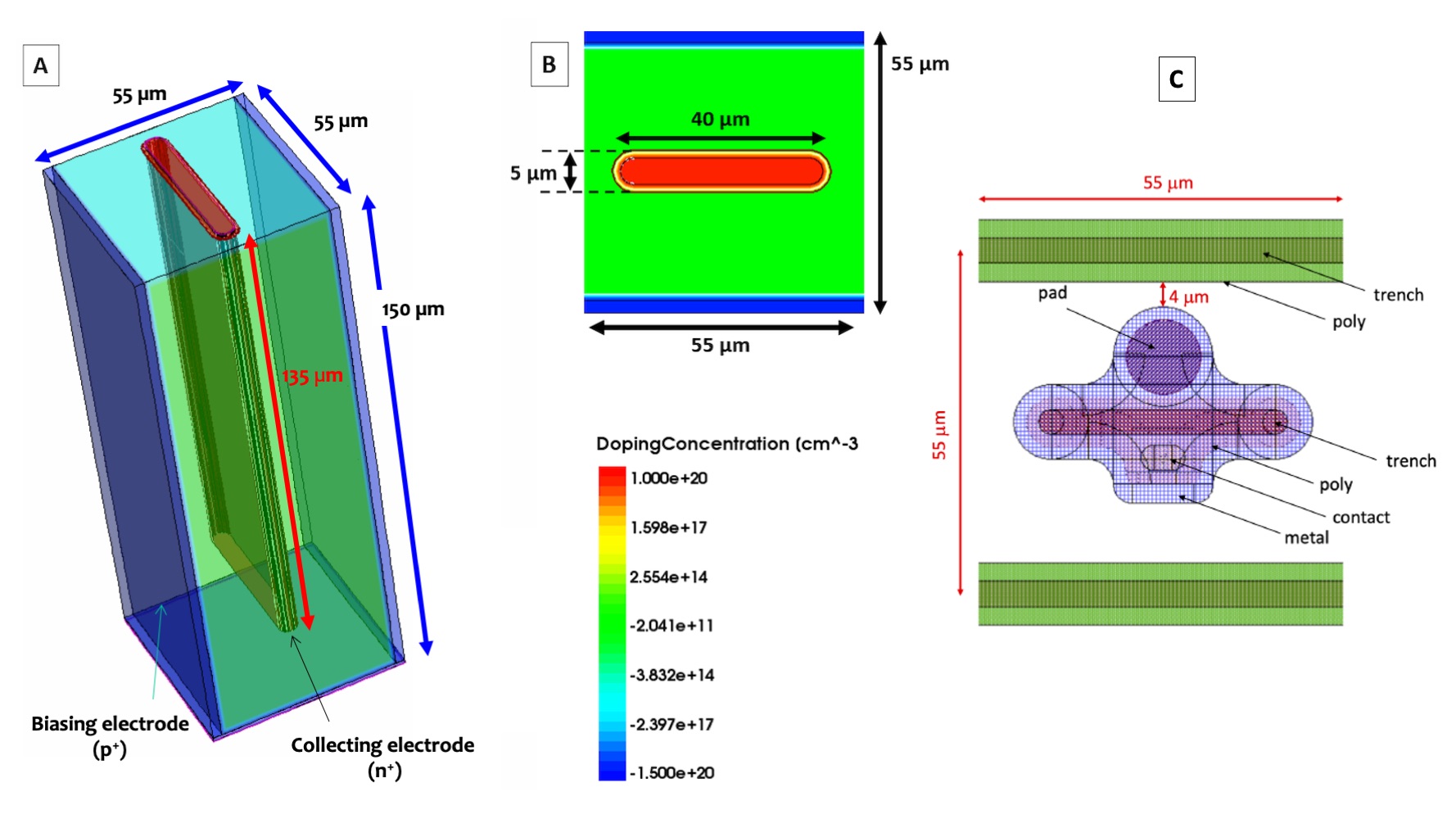}
    \caption{Geometry of the designed 3D-trench pixel, showing dimensions and doping profiles (red for n++ doping, green for p-\,- doping and blue for p++ doping). (A) 3D rendering. (B) Pixel section showing the electrode configuration. (C) Pixel layout.}
    \label{fig:3DPAR}
\end{figure}

The reference elementary cell designed according to the optimised 3D-trench structure is illustrated in figure~\ref{fig:3DPAR}. 
%
Starting from this sensor concept, design optimisation studies led to a defined sizing of the collecting and biasing electrodes. 
While a long collecting trench minimises the weak field area between adjacent trenches, the small distance between them increases the total pixel capacitance, causing a worsening impact on the electronic noise and, as a consequence, on time resolution. The optimised trench size of the collecting electrode has been chosen having \mbox{40~$ \mu$m} in length and \mbox{5~$\mu$m} in thickness, with an inter-electrode distance of  \mbox{15~$\mu$m}. The resulting total pixel capacitance is about $110$~fF.

The sensor has a n-in-p doping profile which guarantees an efficient electron collection at increasing radiation damage~\cite{nINp}. The pixel dimensions (figure~\ref{fig:3DPAR}) were chosen to have a pitch of \mbox{55~$\mu$m} in order to be compatible with the TIMEPIX readout and processing ASIC family~\cite{TIMEPIX}. The electrode configuration presents two external ohmic-wall electrodes which extend over the entire pixel matrix and provide the proper voltage bias to every pixel.
The depth of the sensitive volume is chosen to be \mbox{150~$\mu$m}, as a trade-off between a safe silicon thickness to achieve good uniformity in the shape of the columns during the fabrication process and a sufficient amount of energy deposited by a MIP (about 2~fC). 
The collecting  electrode is \mbox{135~$\mu$m} deep.

For test and complete characterisation of the pixel properties, a variety of different trench-shaped geometries were produced on the wafer, with slightly different trench lengths and widths. In addition, different electrode grouping and read-out configurations were fabricated on the same wafer. A description of the set of produced geometries can be found in refs.~\cite{ForcolinIEEE, ForcolinJINST}.

\begin{figure}[t!]
    \centering
    \includegraphics[width=0.8\textwidth]{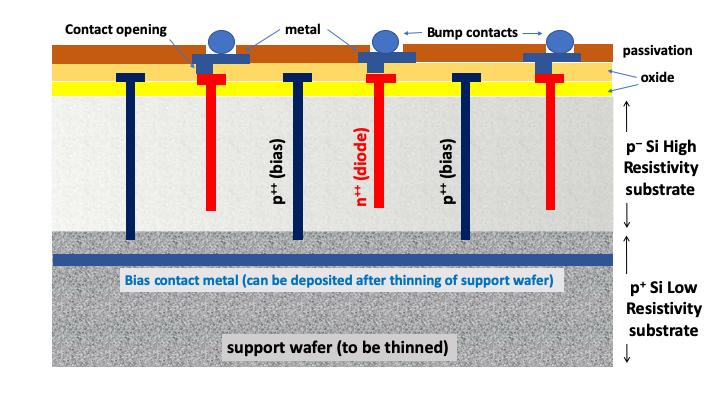}
    \caption{Structure of a 3D-trench sensor (not on scale). The sensitive volume (p Si substrate) has a thickness of \mbox{150~$\mu$m}  and a resistivity of about 5k~$\Omega\,$cm. The support wafer has a thickness of $500 \mu$m and has very low resistivity (some $\Omega\,$cm). The support wafer is usually, but not necessarily, thinned out.}
    \label{fig:3DtrenchSchematic}
\end{figure}

\begin{figure}[h!]
    \centering
    \includegraphics[width=0.8\textwidth,height=0.5\textwidth]{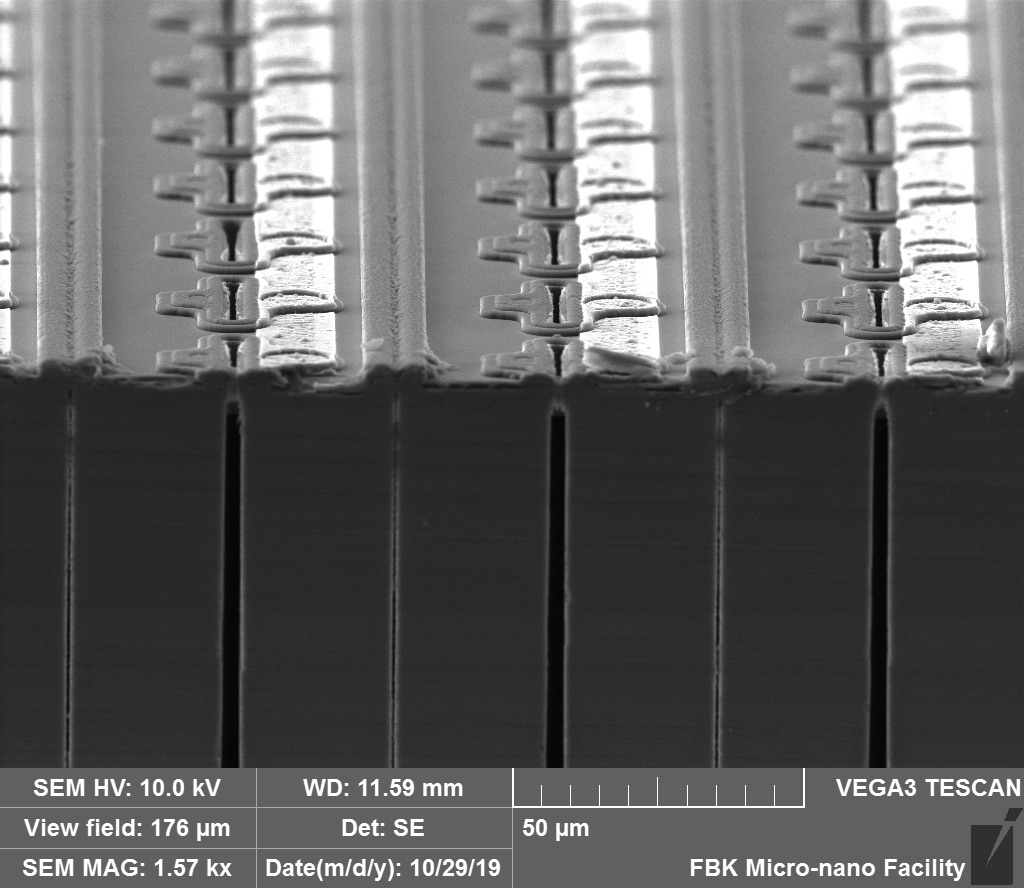}
    \caption{Scanning Electron Microscope picture of a section of the 3D-trench sensor. In this specific case, the read-out is organised in strips (a row of 10 pixels is short-circuited to be read-out by a single front-end channel). The light grey strip is the metal connecting together the different diode (charge collecting) electrodes.}
    \label{fig:3DtrenchSEM}
\end{figure}

 A Single-Sided (Si-Si) process with a support silicon wafer was used for the sensor fabrication. Figure~\ref{fig:3DtrenchSchematic} shows the schematic of the structure of the 3D-trench silicon sensor. The sensors' batch was fabricated at the FBK foundry (Fondazione Bruno Kessler, Trento, Italy). 3D electrodes were made using the Deep Reactive Ion Etching (DRIE) MEMS technique (Bosch process~\cite{Bosch}).
 This technique allows achieving trenches with high aspect ratio (30:1) and good dimensional uniformity. Figure~\ref{fig:3DtrenchSEM} shows a scanning electron microscope (SEM) view of a section of one of the trench structures produced.

\section{Beam Test Measurements}
\subsection{Beam Test Setup}
\begin{figure}[h!]
    \centering
    \includegraphics[width=0.8\textwidth]{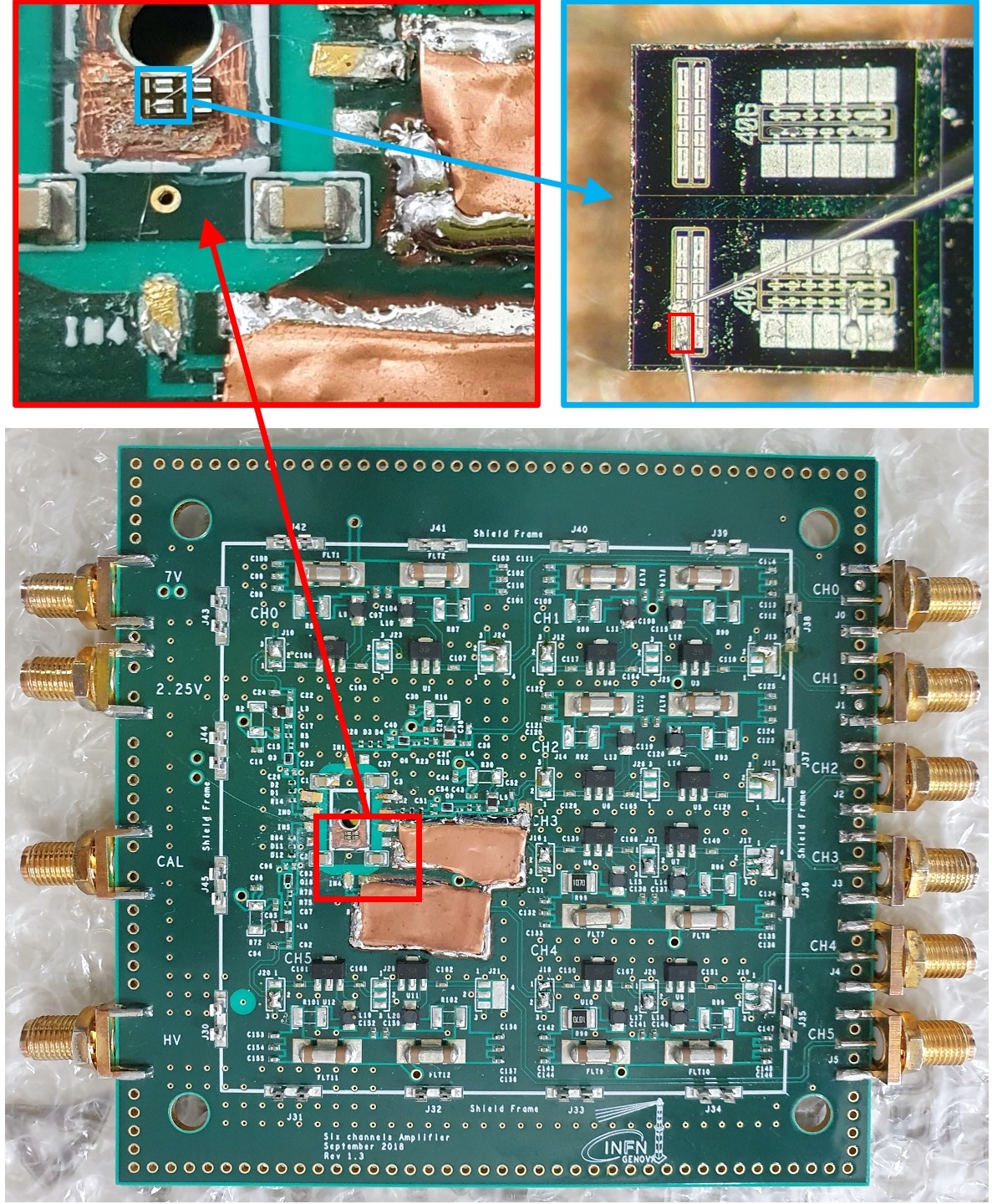}
    \caption{The six-channels front-end board used for the first stage signal amplification. The amplifier connected to the double pixel is located underneath the copper shielding. The 3D-trench silicon sensor used in the tests described in this paper is shown in the top right photograph, and the double pixel which is readout is outlined in red.}
        \label{fig:ampBoard}
\end{figure}

The time resolution of 3D-trench silicon sensors was measured in October 2019 at the PSI $\pi$M1 beamline with 270~MeV/c positive pions, which in silicon produce an energy deposit only slightly larger ($\sim 5$\%) than those from minimum ionising particles (MIPs).
Among the various test structures fabricated, double-pixel structures were tested.
Each structure was attached with conductive tape to a printed circuit boards (PCB) containing the  discrete-components
front-end electronics. 
The sensor output was wire bonded to the input of the amplifier (figure~\ref{fig:ampBoard}).
All sensors were powered from the back contact by supplying a negative high voltage to the pad where they are attached. The 3D sensors were operated at room temperature.

\begin{figure}[h!]
    \centering
    \includegraphics[width=0.65\textwidth]{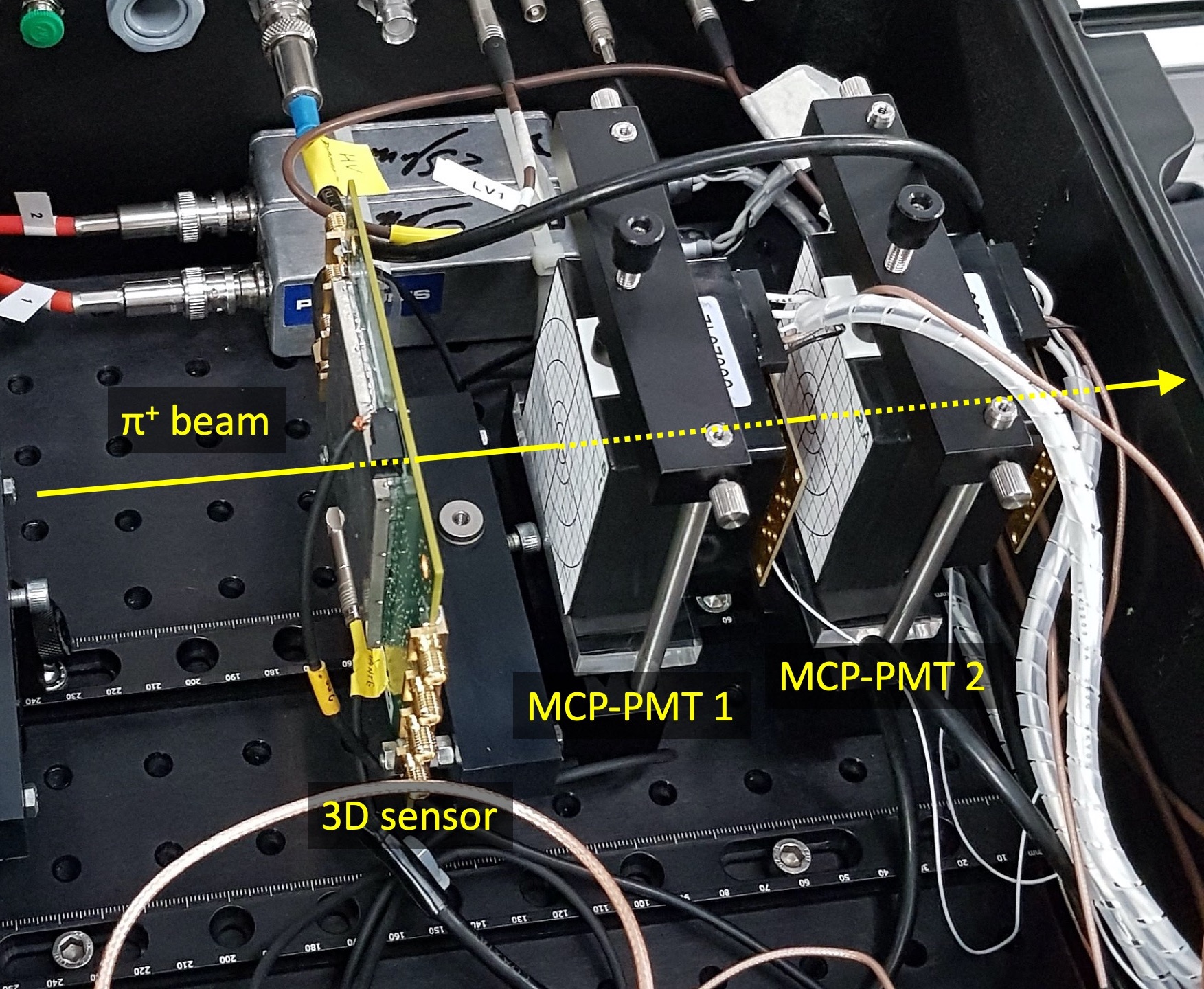}
    \caption{Photograph of the detector setup used during the beam test at PSI. A 3D-trench silicon sensor on the front-end electronic board and the two MCP-PMTs used to provide the time reference are visible in the picture.}
    \label{fig:tbbox}
\end{figure}

The reference measurement of the pion arrival time was provided by two Cherenkov detectors. Each of them consisted of a 20~mm thick quartz radiator~\cite{quartz} attached by means of an optical silicon~\cite{optical-silicon} to a large area (53~mm$\,\times\,$53~mm active window) micro-channel plate photomultiplier tubes (MCP-PMT)~\cite{planacon}.
Both the MCP-PMTs and the PCB with the 3D-trench silicon sensors and front-end electronics were mounted inside a light-tight enclosure positioned on the pion beamline (figure~\ref{fig:tbbox}). 
The silicon sensors were located upstream of the two MCP-PMTs and were transversely aligned with each other with a 1~mm accuracy.
The pions crossed all detectors at normal incidence and provided a uniform illumination of the 3D sensor.

\subsection{Front-End Electronics}
The sensor readout is based on a two-stage signal amplification scheme acting as an inverting trans-impedance amplifier, implemented on a custom-made circuit.
The first amplification stage is performed by an AC-coupled silicon-germanium bipolar transistor designed for high bandwidth (up to  $5 ~\mathrm{GHz}$) and low noise applications, featuring
a gain of nearly $30 ~\mathrm{dB}$ at $2 ~\mathrm{GHz}$ and an integrated output noise of \mbox{260~$\mu$V}. 
This design was optimised for sensors with capacitance $\mathcal{O}(10~\mathrm{pF})$, producing signals with charge $\mathcal{O}(10~\mathrm{fC})$ and a rise time of about $500~\mathrm{ps}$: despite that these are rather different values from those considered in this paper, the board has proved to perform satisfactorily on the signals produced by the new sensors.
The PCB design has been optimised for small and fast signals by minimising all parasitic capacitance and inductance sources, choosing very small size surface-mount components, and ground-burying all signal and power lines whenever possible. Protection from external electromagnetic noise is ensured by hermetic metal shields (figure~\ref{fig:ampBoard}).
The second stage consists of a current amplifier, designed for fast signals. It is based on a monolithic wideband amplifier, with $2~\mathrm{GHz}$ bandwidth, and provides a $20~\mathrm{dB}$ gain factor.
\subsection{Data Acquisition}
\begin{figure}[h!]
    \centering
    \includegraphics[width=0.65\textwidth]{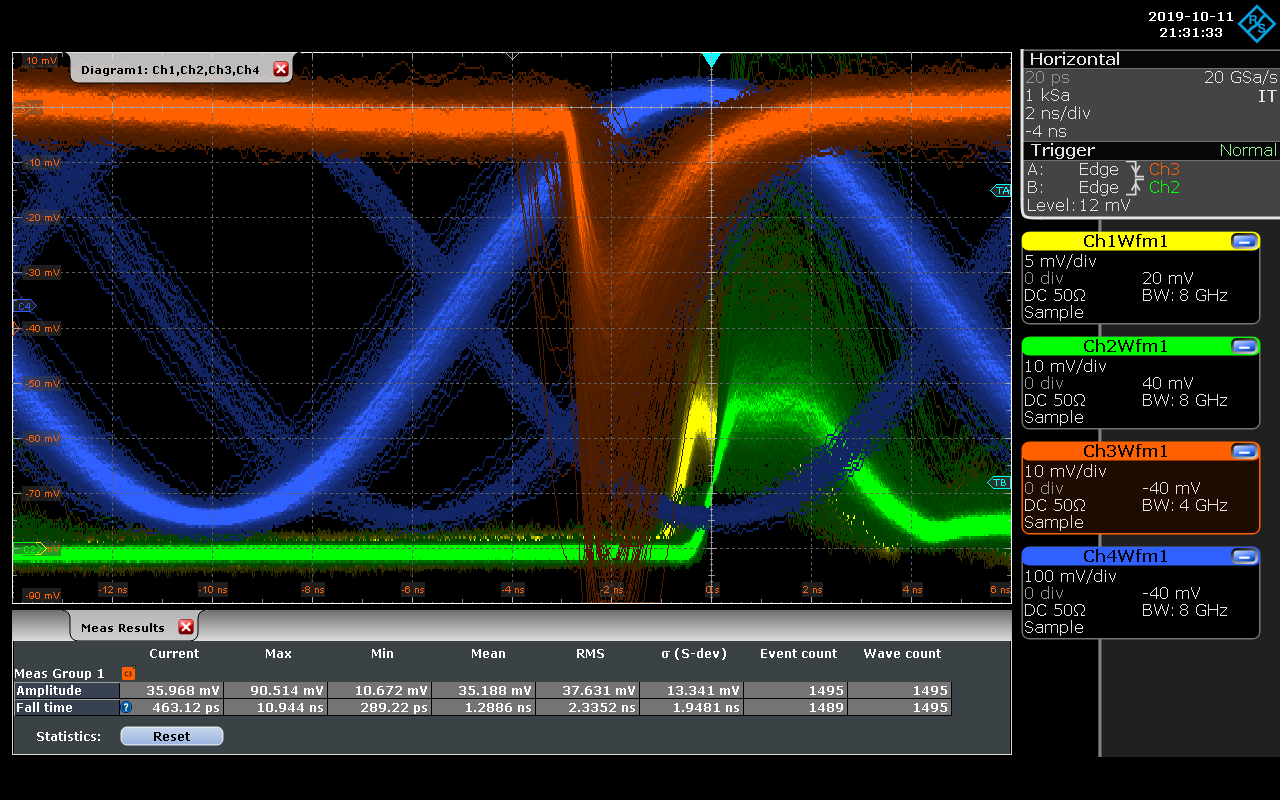}
    \caption{Waveforms acquired with the oscilloscope. The yellow and green waveforms are the reference signals from the MCP-PMTs, the red waveform is the signal from the 3D-trench silicon sensor and the blue waveform is the radio-frequency from the PSI Ring Cyclotron. The three different RF phases correspond to the various components of the beam.}
    \label{fig:oscsignals}
\end{figure}

Signal waveforms from these detectors were acquired by means of a 8~GHz analogue bandwidth, 20~GS/s, 4-channels digital oscilloscope~\cite{oscilloscope} (figure~\ref{fig:oscsignals}). 
The sensor and the MCP-PMTs were connected to the oscilloscope by means of 10~m-long low-loss coaxial cables~\cite{llcable}.
The oscilloscope trigger condition required a signal from the 3D-trench silicon sensor in coincidence with signals from both the MCP-PMTs. 
Trigger thresholds on signals were adjusted to allow an efficient noise rejection while keeping most of the events.

At 270~MeV/c the PSI $\pi$M1 positive beam has a momentum resolution of 0.1\,\% and a transverse size of approximately 40~mm$\,\times\,$40~mm full-width at half-maximum (FWHM) at the focal point where the sensor was located. The beam is mostly composed by pions, with a small contamination of positive muons, positrons and protons.
Since only a small fraction of the particles crossed both the 3D-trench silicon sensor and the two MCP-PMTs, the beam intensity 
was adjusted, by means of collimators, to achieve a data acquisition trigger rate of the order of 100~Hz.
One channel of the oscilloscope was used to record the radio-frequency signal coming from the PSI Ring Cyclotron (RF) and was used to further improve the pion beam purity by selecting a proper delay between the MCP-PMTs signals and the phase of the RF, implementing an effective Time-of-Flight (TOF) detector (figure~\ref{fig:oscsignals}).
The 3D-trench silicon sensor structures tested were double pixels connected together: information like pixel geometrical efficiency or charge sharing between adjacent pixels could not be measured directly.
Samples of 20\,000 events (3\,000 at $V_{\rm bias} = -80$~V) were recorded for different sensor reverse bias and oscilloscope trigger thresholds.

\section{Data Analysis}
\subsection{Method}\label{sec:methods}
\begin{figure}[t!]
    \centering
    \includegraphics[width=0.45\textwidth]{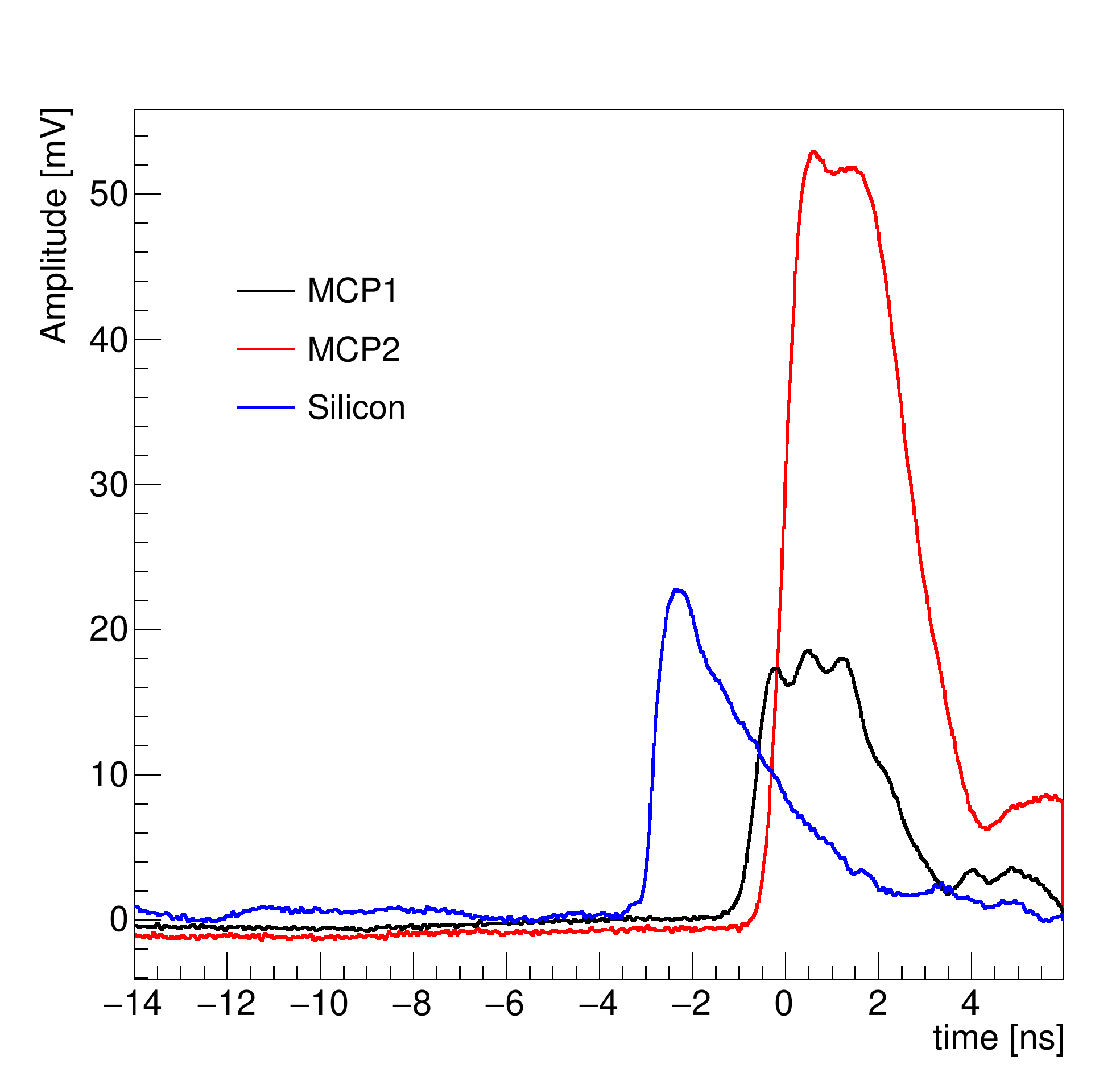}
    \caption{Average waveforms of the (black) MCP-PMT1, (red) MCP-PMT2 and (blue) silicon sensor. In this plot the 3D-trench silicon sensor signal is reversed for convenience. The data points correspond to the average of fifty signals.}
    \label{fig:waveform}
\end{figure}

\begin{figure}[t!]
    \centering
    \includegraphics[width=0.45\textwidth]{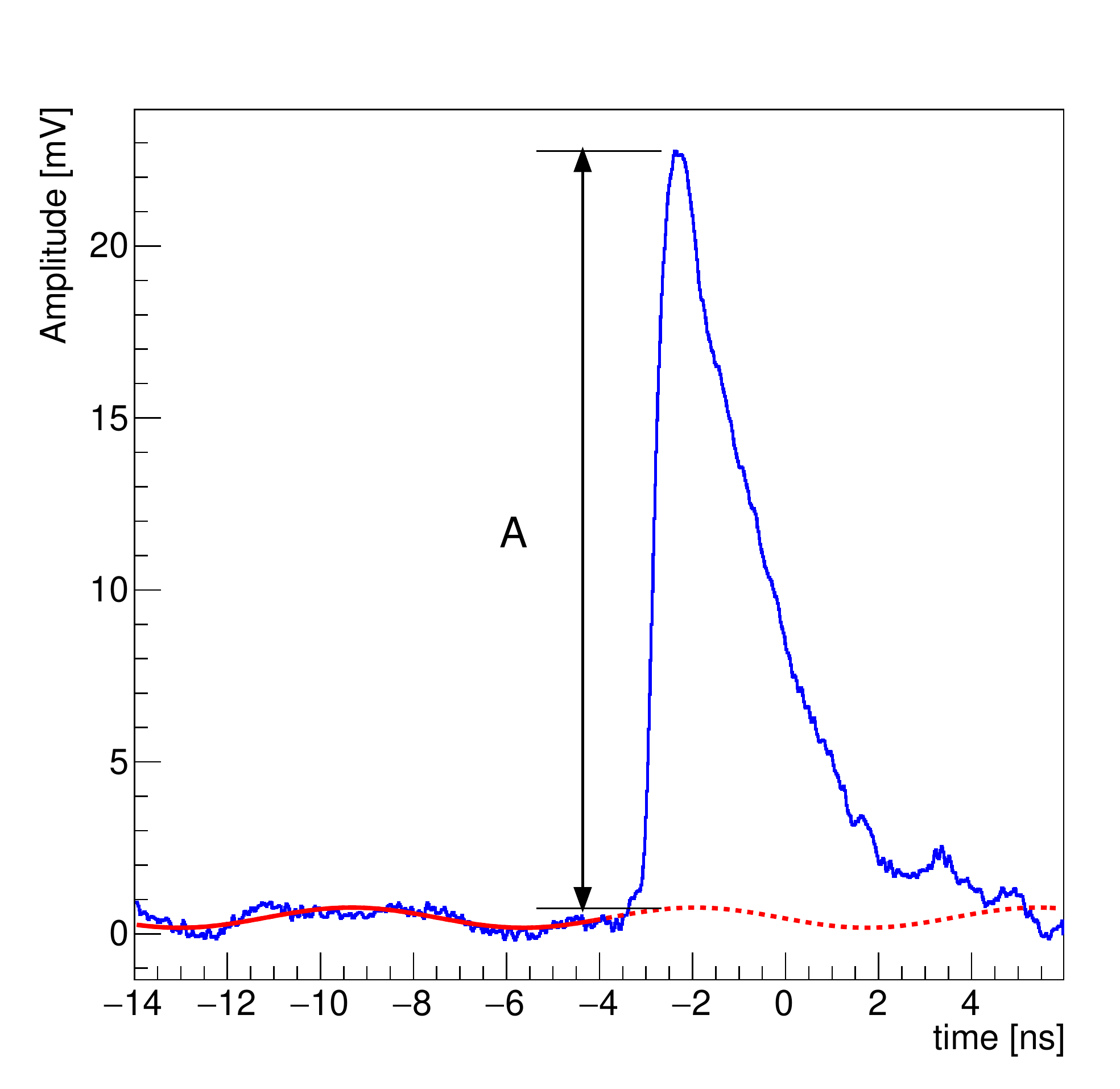}
    \includegraphics[width=0.45\textwidth]{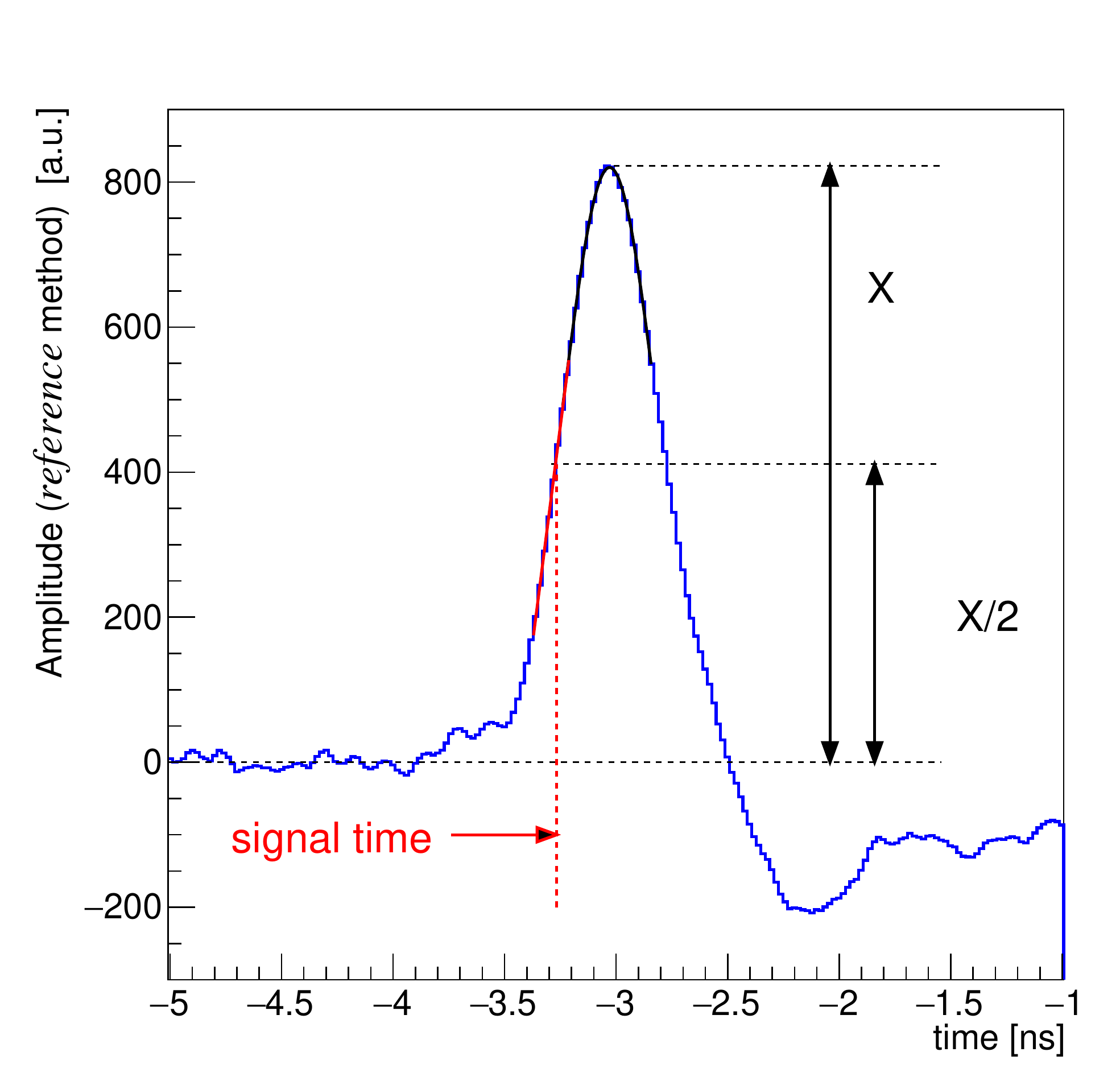}
    \caption{(Left) Average 3D-trench silicon sensor waveform and (right) resulting waveform after the \emph{reference} method is applied. Arrows and functions illustrate how the signal amplitude and the time are determined.}
    \label{fig:CFD2Method}
\end{figure}

Sensor waveforms were analysed with the main purpose of determining the time resolution of the 3D-trench silicon sensor. 
To take into account the attenuation of the 10~m low-loss coaxial cable used during the data acquisition, the cable's transfer function, measured in laboratory, was deconvoluted from the 3D-trench silicon sensor's waveforms.
Figure~\ref{fig:waveform} shows the average shape of the two MCP-PMTs and the 3D-trench silicon sensor waveforms.
The typical rise time values (20-80\% of the signal) are 370, 490 and 200~ps, respectively.
The signal amplitude, $A$, is given by the maximum value of the waveform, corrected by evaluating the baseline just before the beginning of the sensor signal (figure~\ref{fig:CFD2Method} left).
The time of each sensor signal is determined by a method, referred as \emph{reference} in the following, in which from each waveform an identical contribution delayed  by about half of the signal's rise time is subtracted. 
This procedure is crucial for the reduction of the noise and the suppression of the low-frequency background present in the signal waveform.
The resulting waveform, showing a peaking structure, is fitted with a Gaussian function to determine the amplitude (figure~\ref{fig:CFD2Method} right, black line). The time of each waveform is set as the value corresponding to 50\% of the Gaussian's amplitude, by linearly interpolating the signal rising edge (figure~\ref{fig:CFD2Method} right, red line). 
\subsection{Results}
\begin{figure}[t!]
    \centering
    \includegraphics[width=0.45\textwidth]{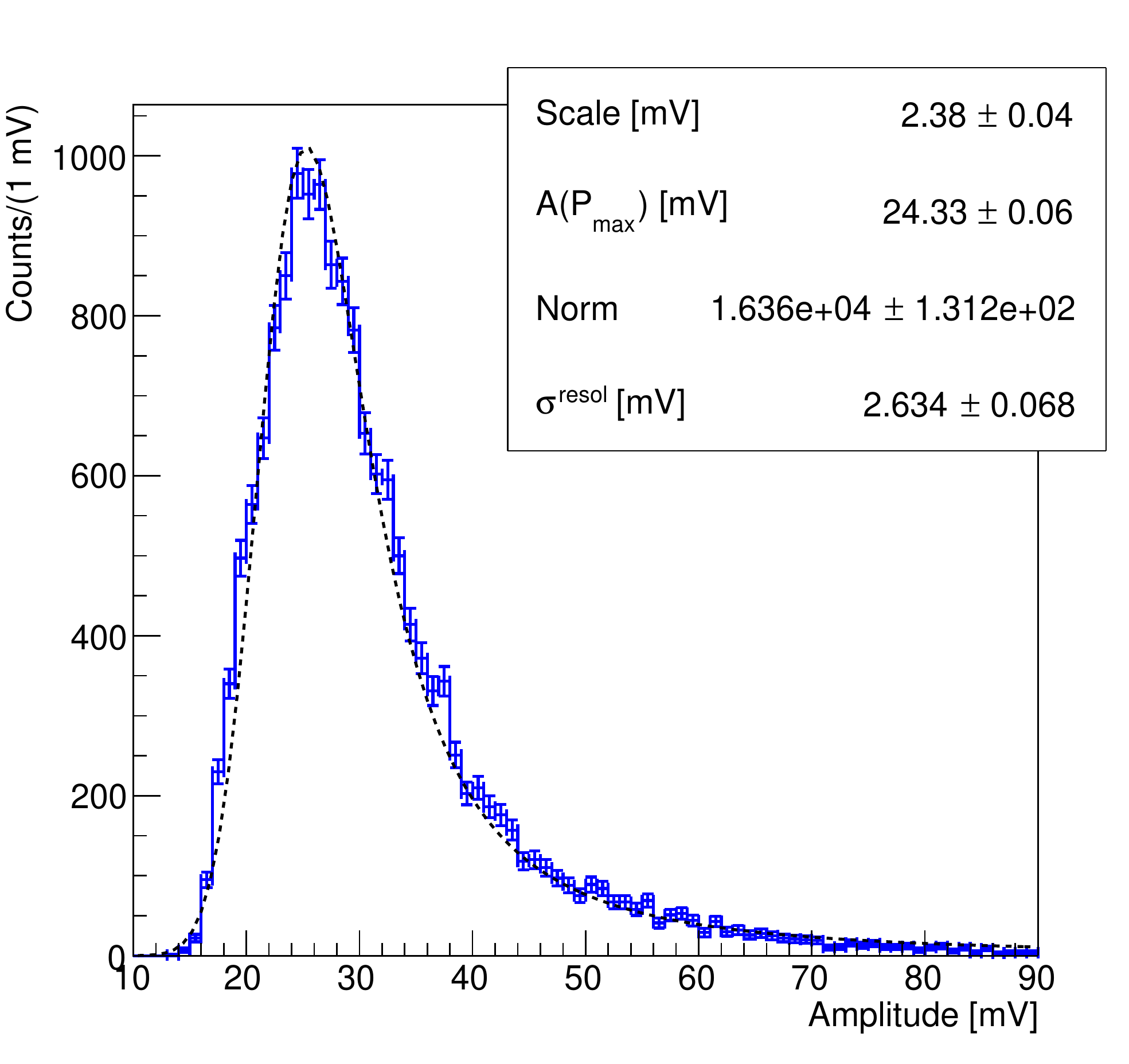}
    \caption{Distribution of the signal amplitudes for the silicon sensor.
    The superimposed blue curve is the result of a fit to a Landau distribution convoluted with a Gaussian. 
    Fit parameters are shown in the legends. }
    \label{fig:Landau}
\end{figure}

The signal amplitude distribution of the silicon sensor corresponding to a sensor bias of $-140$~V is shown in figure~\ref{fig:Landau}. It follows a Landau distribution convoluted with a Gaussian down to the smallest amplitudes, indicating that the trigger threshold does not bias the amplitude distribution of the minimum ionising particles signals.

%
Moreover, the most probable value and the width of the Landau scale as expected for the energy deposit of a MIP in $150 \mu$m of silicon~\cite{Meroli_Landau}, providing an important cross-check of the proper operation of the 3D-trench silicon sensor.

\begin{figure}[h!]
    \centering
    \includegraphics[width=0.45\textwidth]{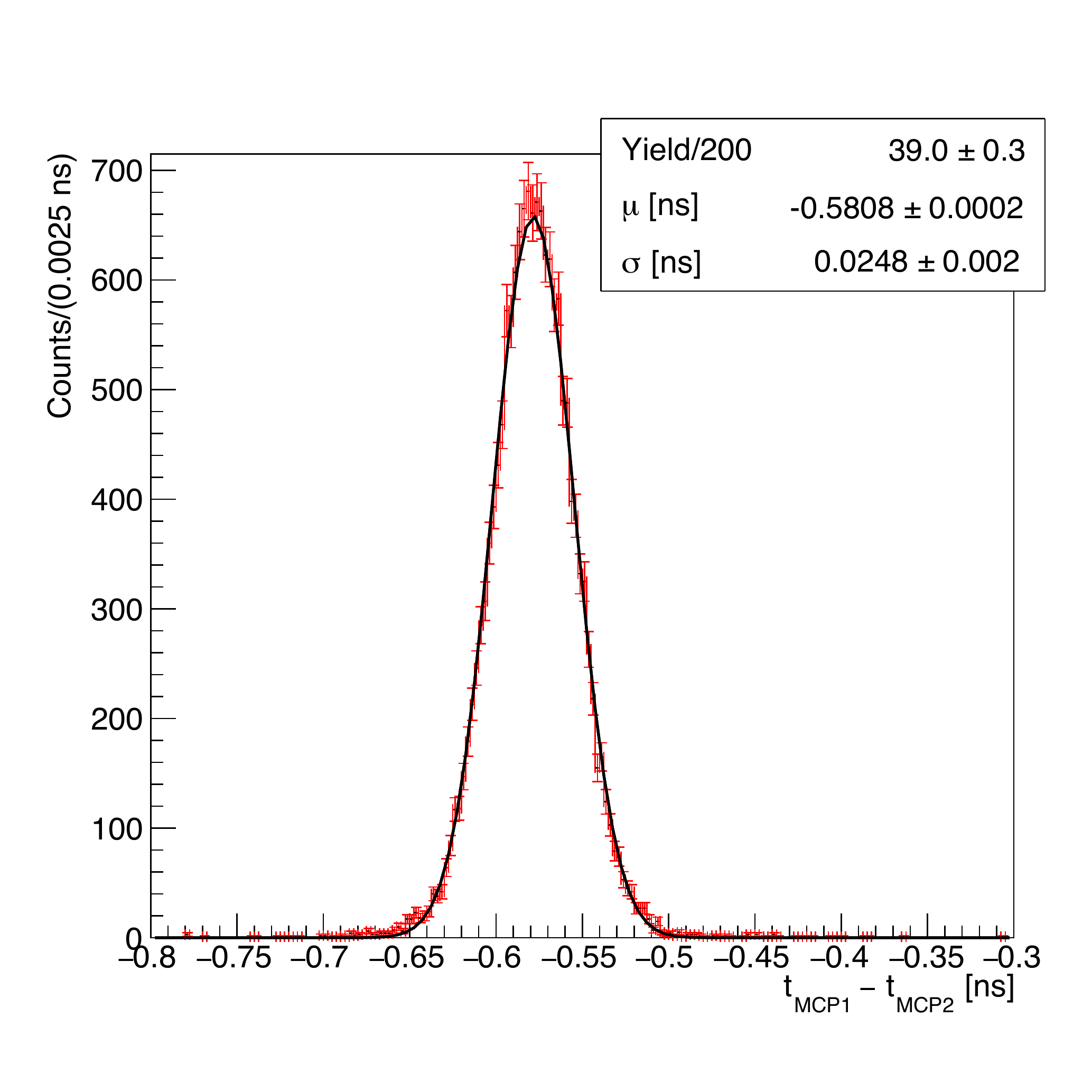}
    \includegraphics[width=0.45\textwidth]{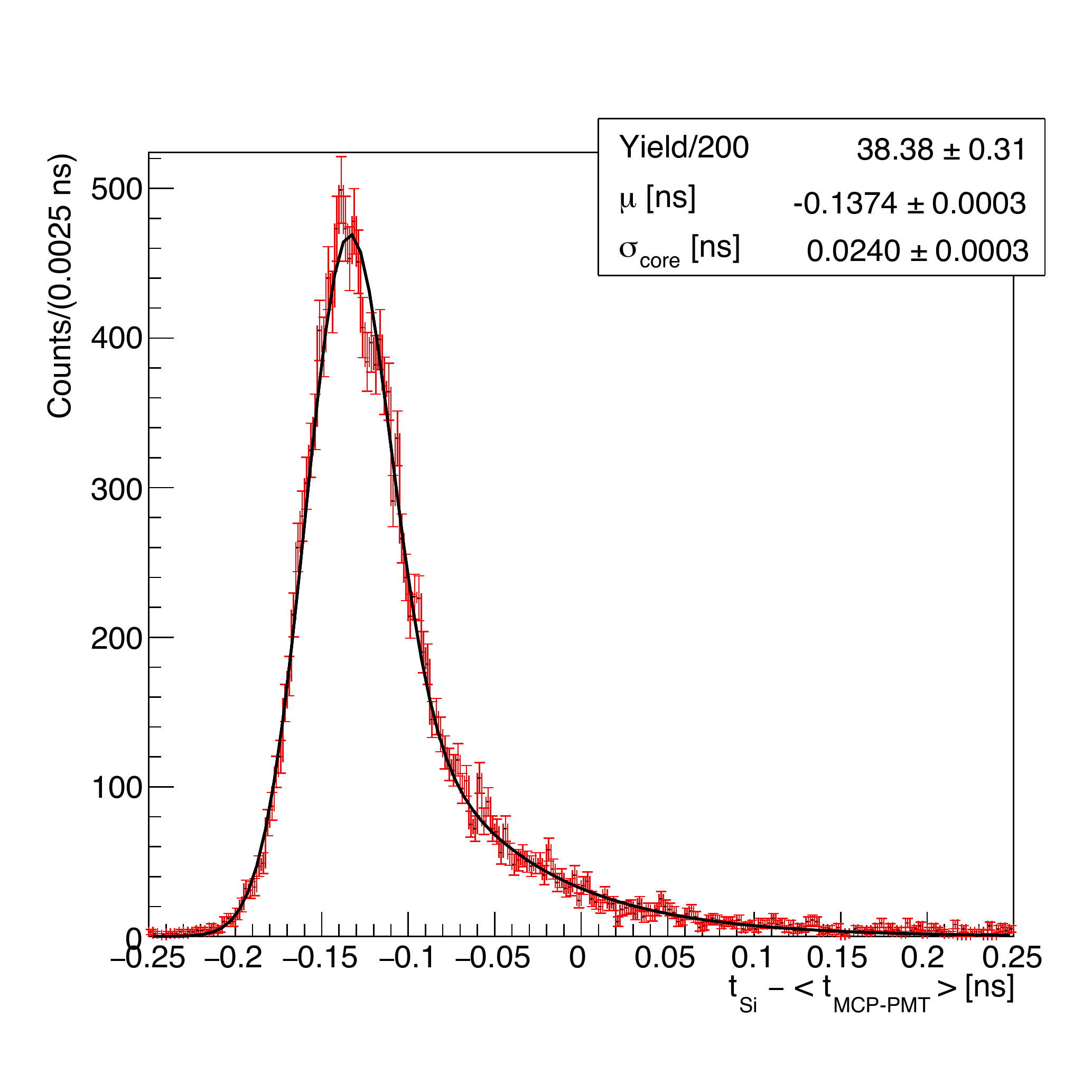}
    \caption{(Left) Distribution of the time difference between the two MCP-PMTs with a Gaussian fit overlaid. (Right) Distribution of the time difference between the 3D-trench silicon sensor and the pion arrival time with the result of the fit overlaid.}
    \label{fig:SummaryResults_timeResolution}
\end{figure}

The time resolution of the 3D-trench silicon sensor was evaluated from the waveform analysis. The delay of the sensor's signal with respect to the pion arrival time was measured using the \emph{reference} method detailed in section~\ref{sec:methods}.
The pion arrival time was given by the average time of the two MCP-PMTs signals, $\langle t_{\rm MCP-PMT}\rangle$. Its accuracy is computed from the width of the distribution of their time difference,
$t_{\rm MCP-PMT1}-t_{\rm MCP-PMT2}$, (figure~\ref{fig:SummaryResults_timeResolution}, left), considering similar resolutions of the two MCP-PMTs.
A time uncertainty (sigma) of $24.8\pm 0.2$~ps is obtained from a Gaussian fit to the data, resulting in a pion arrival time accuracy of about 12.5~ps.

Figure~\ref{fig:SummaryResults_timeResolution} (right) shows the distribution of the time difference between the 3D-trench silicon sensor signal and the pion arrival time, $t_{\rm Si}-\langle t_{\rm MCP-PMT}\rangle$. 
The distribution has a dominant peaking structure with a Gaussian core of $\sigma_{\rm core}=24.0 \pm 0.3$~ps and an exponential tail
of late signals, as expected from simulations (figure~\ref{fig:3Dcomparison}), and time-stamping algorithm effects.
Assuming that the Gaussian core provides an estimate of the sensor performance and
combining it with the pion arrival time uncertainty, the time resolution for the 3D-trench silicon sensor is $\sigma_{\rm t}^{\rm Si} =20.6 \pm 0.4$~ps. 
This measurement is representative of the properties of the full 3D sensor, since the pion beam uniformly illuminates the sensor active area.

\begin{figure}[h!]
    \centering
    \includegraphics[width=0.45\textwidth]{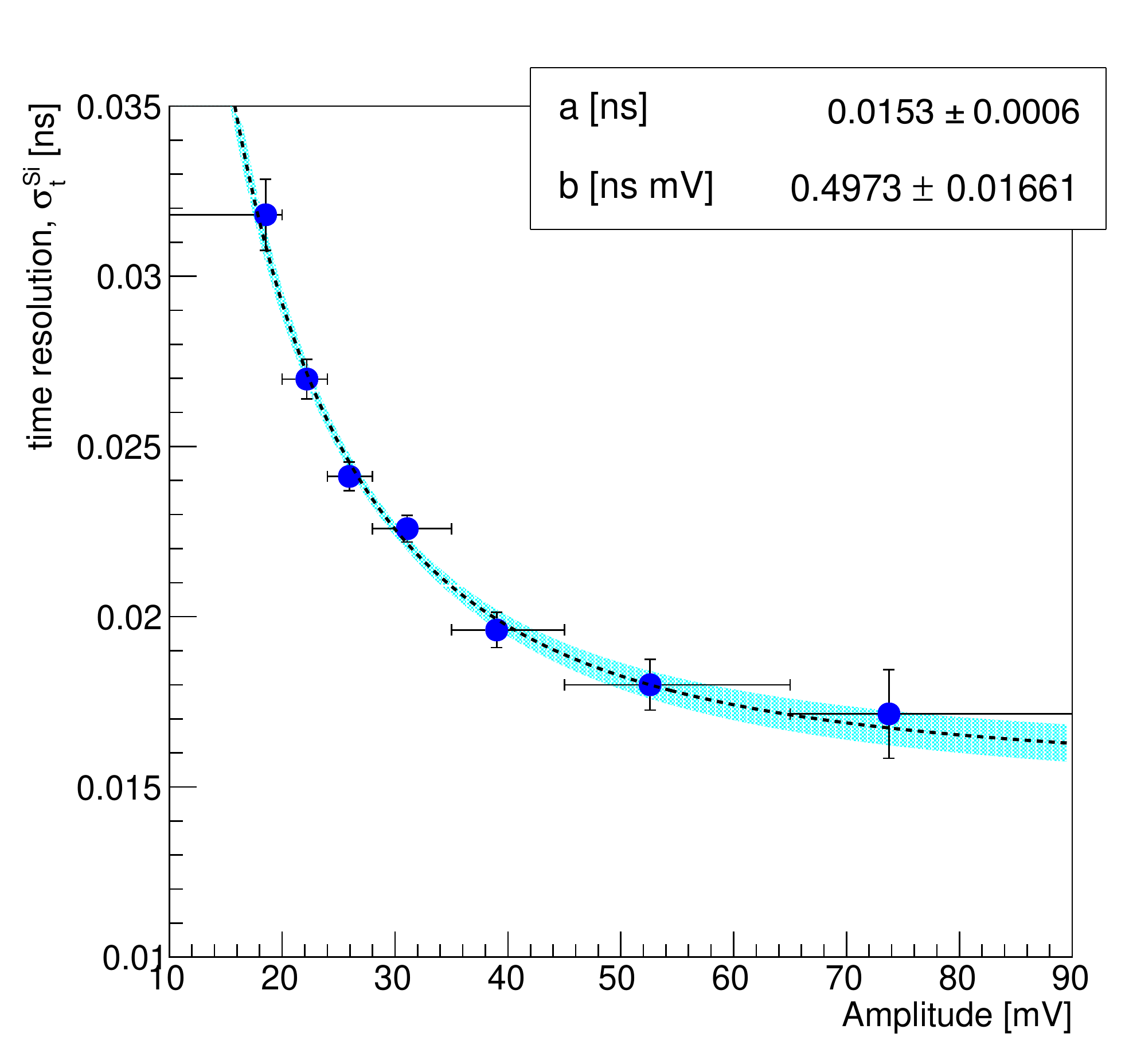}
    \caption{Time resolution of the silicon sensor, $\sigma_{\rm t}^{\rm Si}$, as a function of the signal amplitude fitted to the sum in quadrature of a hyperbola and a constant. Fit parameters are shown in the legends.}
    \label{fig:TimeResolutionVSamp}
\end{figure}

The 3D-trench silicon sensor time resolution varies with the signal amplitude as shown in figure~\ref{fig:TimeResolutionVSamp}, where data points are fitted to the sum in quadrature of a hyperbola and a constant
\begin{equation}\label{eq:FitTimeResolutionVSAmp}
    \sigma_{\rm t}(A) = \sqrt{(a^2 + b^2 / A^2 )}\, ,
\end{equation} 
where $A$ is the signal amplitude and $a$ and $b$ are free parameters.
The asymptotic value $a = 15.3\pm0.6$~ps could be considered as the intrinsic sensor time resolution ($a \simeq \sigma_{\rm un}$) assuming a perfect and noiseless front-end electronics ($\sigma_{\rm ej} \sim 0$) and that the other contribution outlined in eq.~\ref{eq1} are negligible ($\sigma_{\rm tw} \sim 0$, $\sigma_{\rm TDC} \sim 0$).
The parameter $b$, once divided by the mean signal amplitude, provides an estimate of the electronic jitter, $\sigma_{\rm ej} = 17.5\pm 0.6$~ps.
These results indicate that the contribution of the electronic jitter to the time resolution of the sensor is dominant.

The results of the analysis of the data samples acquired at different sensor bias voltages during the test beam are reported in table~\ref{tab:TimeResolution}. 
A rough estimate of the contribution of the electronic jitter  to the time resolution can be computed from the average slew rate and noise as $\sigma_{\rm ej} \sim {\rm N/(dV/dt)}$.
Using the values in table~\ref{tab:TimeResolution} the resulting $\sigma_{\rm ej}$ ranges from 18 to 20~ps confirming that the contribution due to the electronics noise is still the dominant part of the measured time resolution.

\begin{table}[t!]
\centering
    \caption{Average signal-to-noise ratio, noise, slew rate (dV/dt) and time resolution of the 3D-trench silicon sensor for different values of the bias voltage and for different analysis methods.
    The values of the time resolution are subtracted by the pion time of arrival uncertainty.  All results correspond to samples of 20\,000 events, except for that at $V_{\rm bias} = -80$~V, that contains 3\,000 events.}
    \label{tab:TimeResolution}
    \vspace{0.3cm}
\resizebox{1\columnwidth}{!}{
\begin{tabular}{|c| c c c c | c c c c | c |}
\hline
         & \multicolumn{9}{c|}{ method }\\
\cline{2-10}
 setting &\multicolumn{4}{c|}{ \emph{reference} } & \multicolumn{4}{c|}{ \emph{PSI} } & \emph{leading edge}  \\  \hline
 $V_{\rm bias}$ & S/N & N & dV/dt & $\sigma_{\rm t}^{\rm Si}$  & S/N & N  & dV/dt  & $\sigma_{\rm t}^{\rm Si}$  & $\sigma_{\rm t}^{\rm Si}$ \\
 {[V]} & & [mV] & [mV/ps] & [ps] & & [mV] & [mV/ps] & [ps] & [ps]\\ 
\hline
$-20$  
& $12.2$ & $2.22$ & $0.097$ & $24.2 \pm 0.5$ 
& $14.8$ & $2.13$ & $0.070$ & $32.7 \pm 0.7$ & $46.4 \pm 0.5$ \\ 
$-50$  
& $13.0$ & $2.24$ & $0.114$ & $21.9 \pm 0.4$
& $13.1$ & $2.38$ & $0.086$ & $30.3 \pm 0.4$ & $37.6 \pm 0.3$ \\  
$-80$  
& $13.3$ & $2.26$ & $0.121$ & $22.7 \pm 1.2$ 
& $12.2$ & $2.56$ & $0.095$ & $30.0 \pm 1.1$ & $34.2 \pm 1.0$\\ 
$-110$ 
& $13.6$ & $2.26$ & $0.125$ & $20.9 \pm 0.4$  
& $12.3$ & $2.57$ & $0.098$ & $27.8 \pm 0.4$ & $34.7 \pm 0.3$ \\ 
$-140$ 
& $13.9$ & $2.25$ & $0.128$ & $20.6 \pm 0.4$ 
& $12.6$ & $2.56$ & $0.100$ & $27.1 \pm 0.4$ & $35.3 \pm 0.4$ \\ 
\hline
    \end{tabular}
    }
\end{table}

 A different analysis method, referred to as \emph{PSI}, was initially employed for the results reported in refs.~\cite{Vertex,Siena,Hiroshima}.  
In this method the reference time of each waveform is set as the value corresponding to 35\% of the signal's maximum amplitude and is calculated from a linear interpolation of the signal's 20-80\% rising edge.
The results on the time resolution are reported in table~\ref{tab:TimeResolution};
they are $\sim$20-25\% worse than those of the \emph{reference} method, due to a larger noise and smaller slew rate of the signals. 

Following the simplest possible approach (\emph{leading edge} method), the signal time is defined as the value at which the amplitude exceeds a fixed threshold, 
interpolating the waveform in the range of $\pm$40~ps around the threshold. No time-walk correction is applied to account for the variation of the signal amplitude.
The results on the time resolution are reported in table~\ref{tab:TimeResolution}; despite the simplicity of the \emph{leading edge} method the resulting time resolution quickly reaches 35~ps.

\begin{figure}[h]
    \centering
    \includegraphics[width=0.45\textwidth]{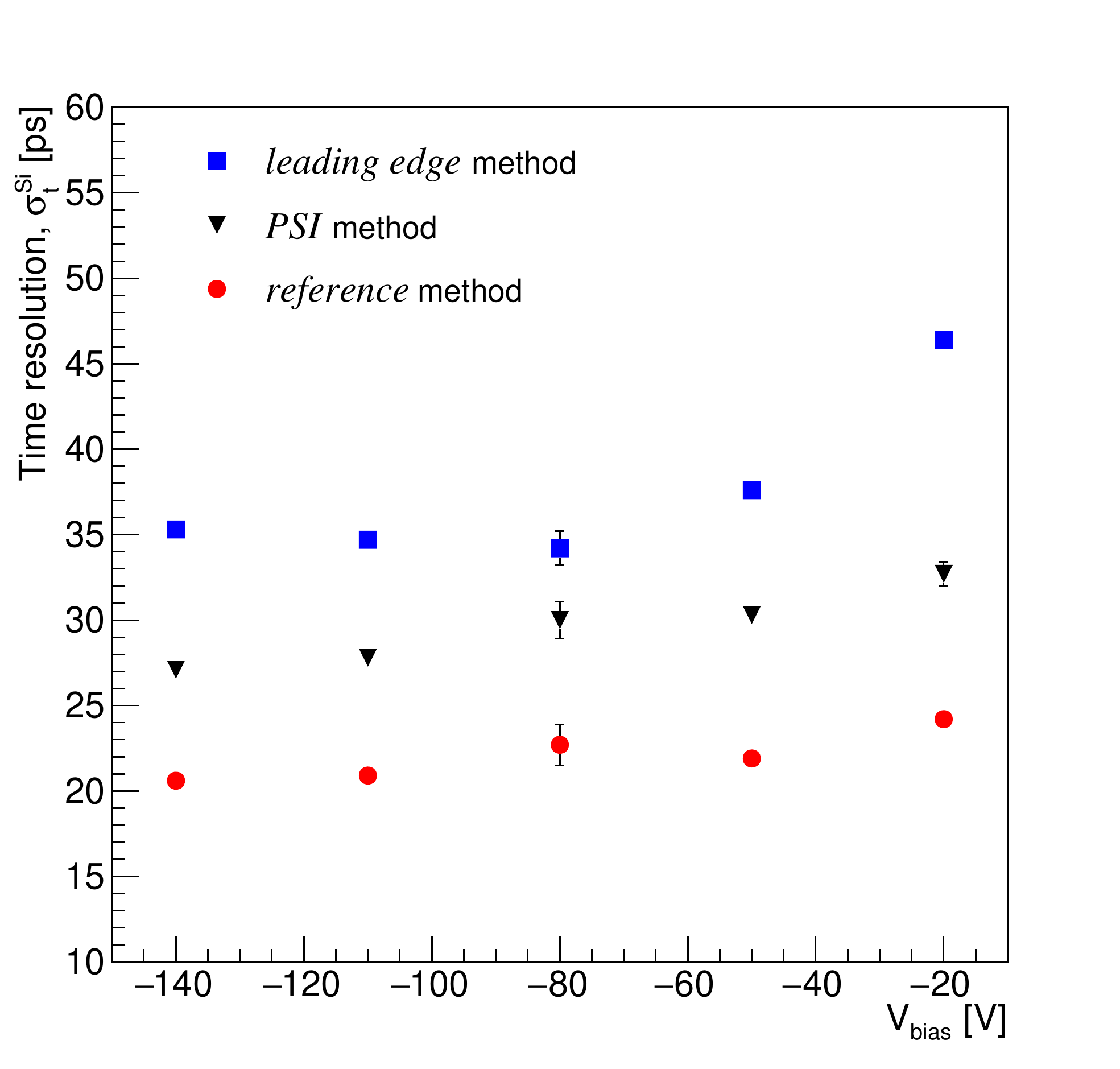}
    \caption{Time resolution of the 3D-trench silicon sensor, $\sigma^{\rm Si}_{\rm t}$, as a function of the sensor bias for different analysis methods considered. The contribution due to the pion time-of-arrival uncertainty is subtracted. 
    }
    \label{fig:ResultsMethods}
\end{figure}

\begin{figure}[t!]
    \centering
    \includegraphics[width=0.45\textwidth]{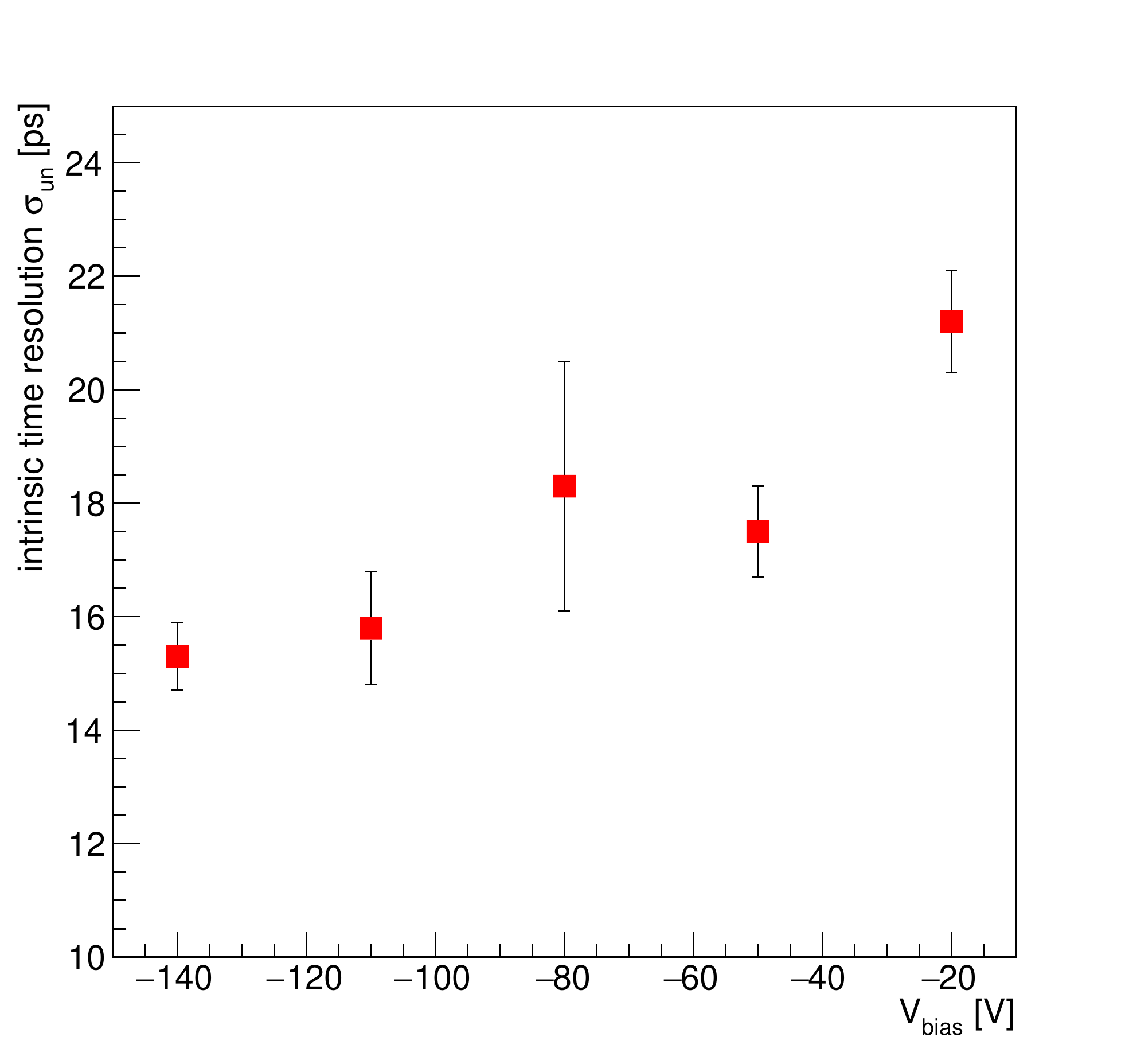}
    \includegraphics[width=0.45\textwidth]{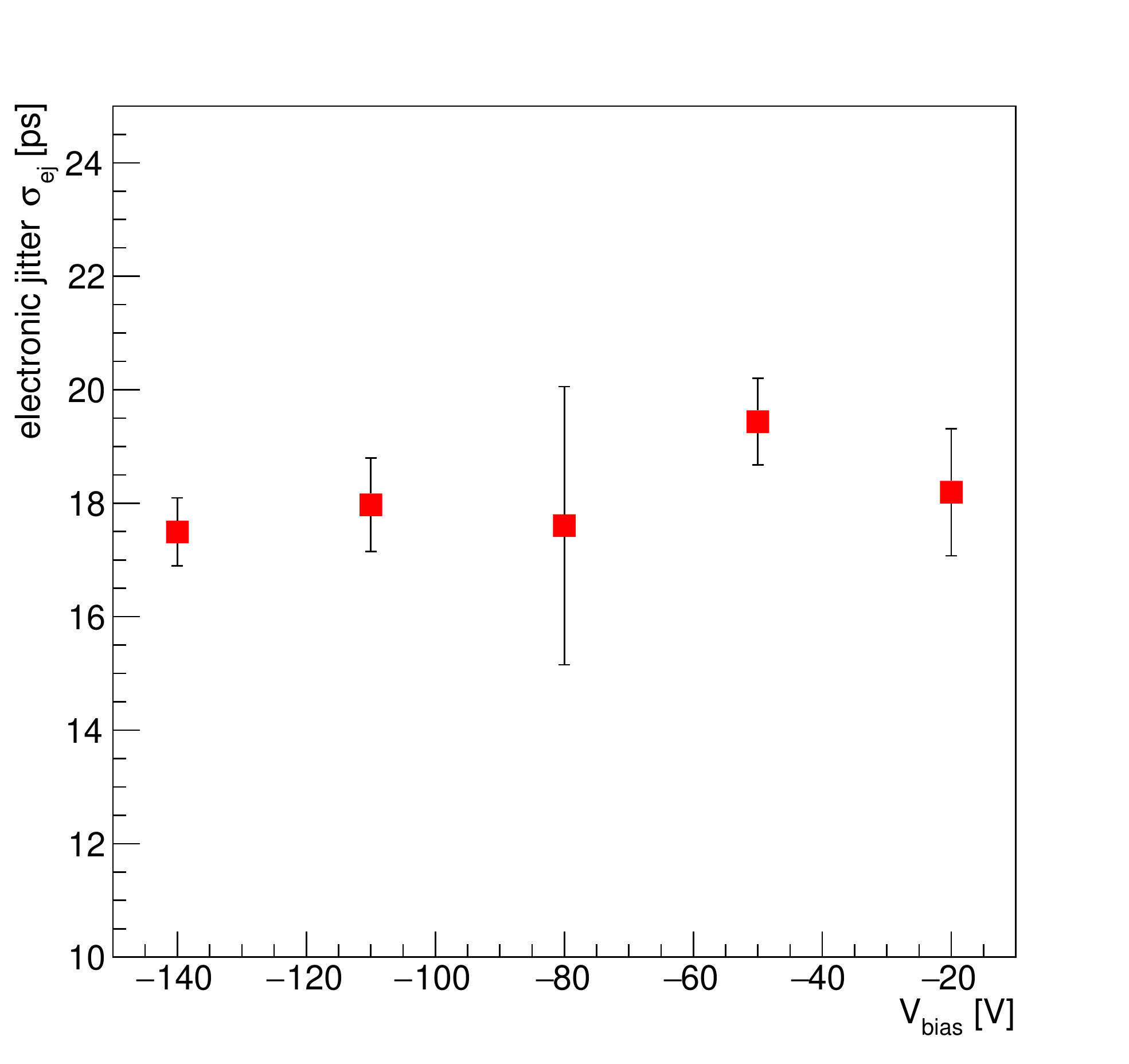}
    \caption{(Left) extrapolated value of the intrinsic time resolution of the 3D-trench silicon sensor and (right) of the electronics jitter determined following eq.~\ref{eq1} for various sensor bias voltages.\label{fig:Parametri_aEb}}
\end{figure}

The results obtained with the various methods and as a function of the detector bias voltage are shown in figure~\ref{fig:ResultsMethods}.
With the exception of the \emph{leading edge} method, a mild dependence on the bias voltage is present.
This dependency is also visible in the values of the intrinsic time resolution shown for the \emph{reference} method in figure~\ref{fig:Parametri_aEb} left, obtained according to the extrapolation described in figure~\ref{fig:TimeResolutionVSamp}. 
As expected, the electronics jitter does not depend on the detector bias voltage, as shown in figure~\ref{fig:Parametri_aEb} right.
On the other hand, the intrinsic time resolution appears to improve while increasing the reverse bias voltage of the sensor.
This behaviour can be explained by the fact that for higher reverse bias voltages the electric field becomes more and more uniform and the charge carrier velocities saturate in most of the detector sensitive volume, thus minimising the $\sigma_{\rm un}$ term in eq.~\ref{eq1}.
As already stated above, the electronics jitter is a relevant contribution to the measured time resolution indicating that an improvement in the front-end electronics is crucial to fully exploit the sensor timing performance.

\section{Conclusions and Outlook}
\label{sec:conclusion}
The 3D-trench silicon sensor described in this paper was designed and produced to address the need to measure with precision the particle arrival time in the next-generation vertex detectors in experiments operating at very high instantaneous luminosities.
This device was designed after thorough simulation studies with the purpose of minimising the variability in charge-collection times among signals generated by particles traversing the sensor active volume.

By using a beam of minimum ionising particles, the time resolution of these new sensors was measured to be about 20~ps at room temperature. 
It was also shown that the measured time resolution contains an important contribution from the electronic jitter of the front-end circuit, indicating that the sensor intrinsic performances have not yet been completely exploited. According to a simplified model discussed in the paper, it is possible to estimate a sensor intrinsic time resolution close to 15~ps.
On the other side, the fast and uniform charge collection time can also be exploited without the need of a sophisticated signal processing procedure. Indeed, it has been shown that a time resolutions of 35~ps at room temperature can be reached by means of the simplest possible measuring procedure, that is a leading-edge discriminator without any time-walk correction.

The results obtained and the clear indications about possible performance improvements place 3D-trench silicon sensor at the cutting edge of the development activities for high-resolution timing sensors suitable for charged particle detection. 
In particular, 3D-trench pixel sensors now stand out as a valuable option to be seriously considered in the conception of tracking detectors of high-energy physics experiments of the next decades. 

\acknowledgments
{
This work was supported by the Fifth Scientific
Commission (CSN5) of the Italian National
Institute for Nuclear Physics (INFN), within the Project TIMESPOT and by the ATTRACT-EU initiative, INSTANT project.
The authors wish to thank the whole technical and scientific staff of Paul Scherrer Institute for their kindness and effectiveness. A special thank is due to Angela Papa and Manuel Schmidt for their valuable help.
Last but not least, the authors are extremely grateful to Rohde\&Schwartz for their kindness in providing, for free and for the entire period of the tests, the high-end oscilloscope used for data acquisition and measurements.
}

\bibliographystyle{ieeetr}  
\bibliography{000AAA-paper}  

\end{document}